\documentclass[aps,prl,twocolumn,notitlepage,showpacs,superscriptaddress,am]{revtex4-2}%
\usepackage{graphicx}
\usepackage{amsmath}
\usepackage{amssymb}
\usepackage{color}
\usepackage{amsfonts}%
\usepackage[breaklinks=true,colorlinks=true,linkcolor=blue,urlcolor=blue,citecolor=blue]{hyperref}

\providecommand{\U}[1]{\protect\rule{.1in}{.1in}}

\def\ET{(BEDT-TTF)$_2X$}

\def\stf{$\kappa$-STF$_x$}
\def\STF{$\kappa$-[(BEDT-STF)$_x$(BEDT-TTF)$_{1-x}$]$\rm _2 Cu_2 (CN)_3$}

\def\Cu{$\kappa$-(BEDT-TTF)$_2$Cu$_2$(CN)$_3$}

\def\aI3{$\alpha$-I$_3$}
\def\iToneT{$(T_1T)^{-1}$}
\def\iTone{$T_1^{-1}$}

\def\TFL{$T_{\rm FL}$}

\def\sro{Sr$_2$RuO$_4$}
\def\sto{SrTiO$_3$}

\renewcommand{\vec}{\boldsymbol}

\usepackage{xcolor}

\usepackage{changes}

\begin{document}
\title{
Quasiparticle to local moment crossover in bad metals
}

\author{A. Chen}
\thanks{These authors contributed equally to this work}
\affiliation{Institute of Solid State Physics, TU Wien, 1040 Vienna, Austria}
\author{F. B. Kugler}
\thanks{These authors contributed equally to this work}
\affiliation{Institute for Theoretical Physics, University of Cologne, 50937 Cologne, Germany}
\affiliation{Center for Computational Quantum Physics, Flatiron Institute, New York 10010, USA}
\author{P. Dole\v{z}al}
\affiliation{Institute of Solid State Physics, TU Wien, 1040 Vienna, Austria}
\affiliation{Department of Condensed Matter Physics, Faculty of Mathematics and Physics,
Charles University, Ke Karlovu 5, 121 16 Prague 2, Czech Republic}
\author{Y. Saito}
\affiliation{Department of Physics, Graduate School of Science, Hokkaido University, Sapporo 060-0810, Japan}
\affiliation{Institute of Physics, Goethe-University Frankfurt, 60438 Frankfurt (Main), Germany}
\author{A. Kawamoto}
\affiliation{Department of Physics, Graduate School of Science, Hokkaido University, Sapporo 060-0810, Japan}
\author{A. Georges}
\affiliation{Collège de France, PSL University, 75005 Paris, France}
\affiliation{Center for Computational Quantum Physics, Flatiron Institute, New York 10010, USA}
\affiliation{Department of Quantum Matter Physics, University of Geneva, 1211 Geneva, Switzerland}
\affiliation{Centre de Physique Théorique, Ecole Polytechnique, 91128 Palaiseau, France}
\author{A. Pustogow}
\email{pustogow@ifp.tuwien.ac.at}
\affiliation{Institute of Solid State Physics, TU Wien, 1040 Vienna, Austria}


\begin{abstract}
Non-Fermi-liquid charge transport in the vicinity of electronic instabilities has been intensely studied for decades. Deviations from  $\rho_{\rm FL}=\rho_0+AT^2$ in \textit{bad} and \textit{strange} metals are commonly ascribed to a breakdown of Landau's quasiparticle (QP) concept. 
Yet, it remains unclear what mechanism drives the temperature dependence of $\rho(T)$ beyond $\rho_{\rm FL}$. 
Here, we examine the bad metal upon approaching the Mott metal-insulator transition via chemical pressure in \STF. Through nuclear magnetic resonance (NMR) and transport experiments on the same single crystals, we directly link the onset of deviations from Korringa law \iToneT $= \mathrm{const.}$ with the rise of $\rho(T)$ beyond $\rho_{\rm FL}$. 
From the NMR relaxation rate, we can identify the gradual crossover between the QP-dominated regime at low $T$ to predominant local moments at higher $T$.
By comparing our experimental findings with dynamical mean-field theory calculations, which accurately reproduce the transport data, 
we reveal how this crossover is reflected in $T$-dependent changes of the QP spectrum.
Near the Mott insulator, where $d\rho/dT<0$ at high $T$, an Einstein-relation analysis shows that bad-metal behavior with $d\rho/dT>0$ is driven by the temperature dependence of the electronic compressibility rather than the diffusion constant.

\end{abstract}

\maketitle


Landau's quasiparticle (QP) picture 
within Fermi-liquid (FL) theory 
serves as 
the cornerstone of today's understanding of charge transport in metals~\cite{Landau1956}. However, one of its main predictions---the rise of direct-current (dc) resistivity $\rho$ quadratic with temperature $T$
as the scattering phase space grows with energy~\cite{Gurzhi1959}---is commonly concealed by other sources of scattering, e.g.\ from impurities or phonons. Electron-electron scattering becomes clearly visible when correlations enhance the effective mass $m^{\star}$ 
close to electronic instabilities, such as quantum critical points (QCPs) and metal-insulator transitions (MITs). 
The hallmark $\rho_{\rm FL}(T)=\rho_0 \!+\! AT^2$ behavior
is obeyed at low enough 
$T \!\lesssim\! T_{\rm FL}$. 
The $A$-coefficient usually obeys 
the Kadowaki--Woods relation $A \!\propto\! \gamma^2 \!\propto\! (m^{\star})^2$, 
with $\gamma$ the Sommerfeld coefficient~\cite{Kadowaki1986}. 
Remarkably, this scaling holds for diverse materials classes, from heavy fermions over transition-metal oxides to organic conductors~\cite{Jacko2009}.

\begin{figure}[h!]
\centering
 \includegraphics[width=0.85\columnwidth]{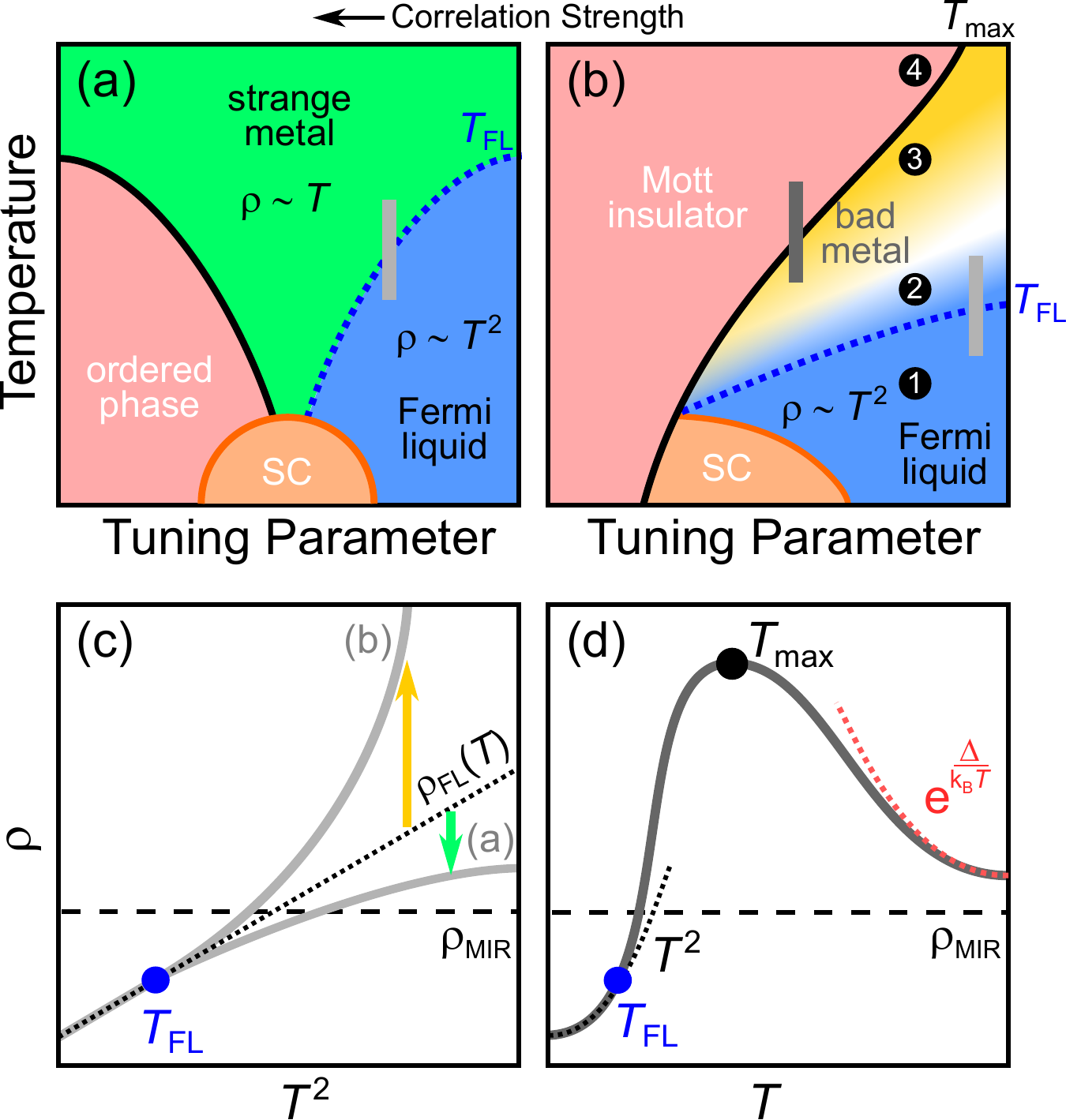}
\caption{Phase diagrams of (a) a strange metal above a QCP and (b) a bad metal nearby a Mott MIT. (c) While $\rho(T) \propto T$ in (a) yields a slower rise than $\rho_{\mathrm{FL}}(T)$, in (b) $\rho(T)$ rises faster than $T^2$~\cite{Kurosaki2005,Shimizu2011,Shimizu2016,Furukawa2018,Pustogow2021-Landau,Pustogow2021-percolation,Pustogow2023,Lin2015}. 
(d) The transition from bad metal to Mott insulator yields a resistivity maximum with $\rho \gg \rho_{\rm MIR}$~\cite{Kurosaki2005,Shimizu2011,Shimizu2016,Furukawa2018,Pustogow2021-Landau,Pustogow2021-percolation,Pustogow2023}.
}
\label{intro}
\end{figure} 

At the same time, transport in correlated electron systems  
behaves very differently at higher $T > T_{\rm FL}$, especially 
if $T_{\rm FL}$ is suppressed near a QCP or MIT~\cite{Imada1998,Lohneysen2007,Keimer2015,Phillips2023}, see Fig.~\ref{intro}. 
While around QCPs strange-metal behavior $\rho\propto T$ can be observed from 
absolute zero up to almost 1000~K~\cite{Takagi1992,Custers2003,Bruin2013,Legros2019,Cao2020,Nguyen2021}, an increase of $\rho$ faster than $T^2$ has been observed for $T>T_{\rm FL}$ in single-band materials such as organic Mott systems~\cite{Kurosaki2005,Shimizu2011,Shimizu2016,Furukawa2018,Pustogow2021-Landau,Pustogow2021-percolation,Pustogow2023} and weakly doped SrTiO$_3$~\cite{Lin2015}.
Note that virtually all of these systems cross over from 
FL to
bad metal with $\rho(T)$ rising well above the Mott--Ioffe--Regel (MIR) limit $\rho_{\rm MIR}$ 
with no
sign of saturation~\cite{Gunnarsson2003,Hussey2004,Takenaka2005}.
How to understand and describe these different regimes of transport in a consistent framework
remains a topic of active discussion~\cite{Deng2013,Hartnoll2015,Keimer2015,Phillips2023}.

In a simple Drude picture, the $T$ dependence of $\rho=\left(Ne^2\tau/m\right)^{-1}$ 
may come from
different factors. 
On the one hand, transport in insulators arises from a thermally activated carrier density $N$ across an energy gap, 
yielding $\rho(T)\propto N^{-1}(T) \propto e^{\Delta/k_{\rm B}T}$. 
On the other hand, in metals, the $T$ dependence of $\rho$ is commonly dominated by that of the scattering rate: $\rho(T) \propto\tau^{-1}(T)$. 
In a typical good metal obeying Landau's FL behavior, 
$\tau^{-1}(T)\propto T^2$ leads to the 
quadratic $T$ dependence of $\rho_{\rm FL}$. 
Note that $m$ and $N$ are often assumed to be $T$-independent in metals due to  
a constant density of states around the Fermi level.
In this framework, low-$T$ metallic transport $d\rho/dT>0$ with deviations from $\rho_{\rm FL}$  
has been discussed as a potential breakdown of FL theory, suggesting fundamentally different mechanisms of charge transport such as Planckian dissipation~\cite{Zaanen2004,Bruin2013,Hartnoll2015,Keimer2015,
Chowdhury2022,Phillips2023}. 

Here, we investigate the transport behavior of $\kappa$-\ET\ organic materials. Close to their Mott MIT, 
these materials display four distinct regimes of transport schematized in Fig.~\ref{intro}(b,d). 
For $T<T_{\rm FL}$,
FL behavior applies; 
for $T>T_{\rm max}$,
insulating-like behavior with $d\rho/dT < 0$ is observed. 
These two regimes are separated by two intermediate regimes: 
(2) with transport due to resilient QPs \cite{Deng2013} and $\rho$ increasing faster 
than $\rho_{\mathrm{FL}}$, 
and (3) a bad metal with $\rho$ exceeding $\rho_{\rm MIR}$. 
Through simultaneous transport and nuclear magnetic resonance (NMR) experiments, we 
reveal a direct connection between 
the breakdown of Korringa law and the deviation of $\rho$ from $\rho_{\rm FL}$ in (2). 
Furthermore, we show that local moments already start forming in 
regime (3) where transport is that of a bad metal with $d\rho/dT > 0$.
Through dynamical mean-field theory (DMFT) calculations~\cite{Georges1996}, we
show that these four regimes of transport can be described within a unique theoretical framework. 
These calculations reveal how QPs start to lose coherence for $T>T_{\rm FL}$ when entering regime (2) 
and eventually fully disappear leaving a thermally filled pseudogap in the excitation spectrum for $T>T_{\rm max}$ 
in regime (4).
Furthermore, we show that metallic transport with $\rho \neq \rho_{\rm FL}$ at
$T \gtrsim T_{\rm FL}$ can be understood by generalizing the QP concept with energy- and/or $T$-dependent QP properties \cite{prange_1964,Deng2013,Georges2021}. 
The resistivity maximum and regime (4) are best understood however by adopting an 
Einstein relation description of transport.

\begin{figure}
\centering
 \includegraphics[width=1\columnwidth]{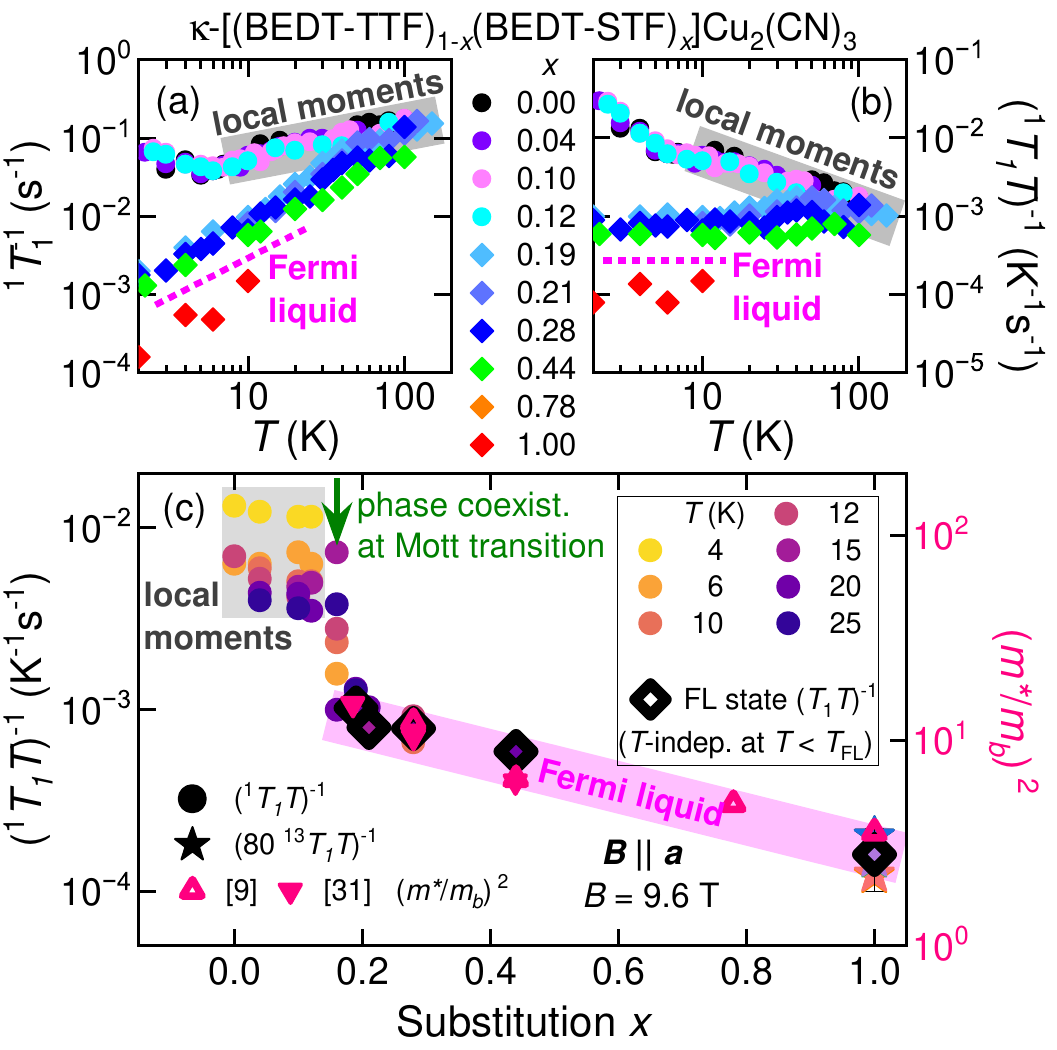}
\caption{(a) The Mott transition in \stf\ yields a bifurcation of insulating and metallic NMR relaxation. (b) The pronounced $T$-dependence and generally faster \iToneT\ for $x\leq 0.16$ are consistent with the response of local moments. In the low-$T$ Fermi-liquid state ($x\geq 0.19$) Korringa-type behavior with $T$-independent \iToneT\ is established. (c) By comparing our $^1$H NMR results to optical and specific heat data~\cite{Pustogow2021-Landau,Yesil2023}, we find that \iToneT\ scales with the effective mass enhancement $\left( m^{\star}/m \right)^2$.
}
\label{T1T-MIT}
\end{figure}

\begin{figure*}[ptb]
\centering
\includegraphics[width=2.05\columnwidth]{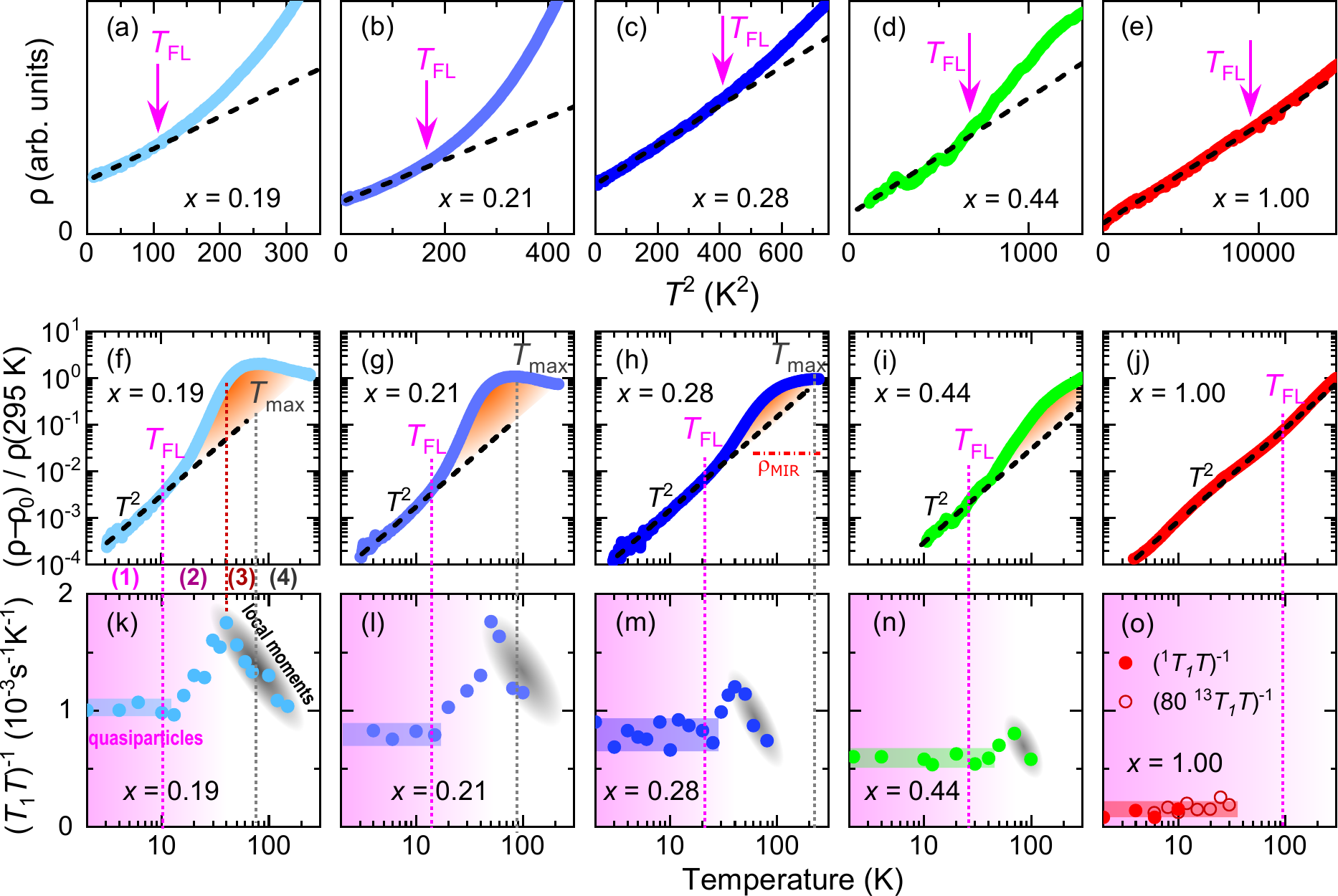}
\caption{Combined assessment of FL and bad-metal behavior in resistivity and NMR relaxation. The plots versus $T^2$ (a-e) and on double-logarithmic scales (f-j) reveal $\rho(T)=\rho_0+AT^2$ at $T< $ \TFL. Above that, the increase is steeper than quadratic (orange shaded) until $\rho$ reaches a maximum at $T_{\rm max}$ \cite{Pustogow2021-Landau,Pustogow2021-percolation}. Sample geometry precluded precise determination of absolute resistivity, see End Matter. (k-o) \iToneT\ probed by $^1$H NMR (in panel (o) appropriately scaled $^{13}$C data were added) is (1) $T$-independent at $T\leq $ \TFL\ and (2) increases at higher $T$ 
where $\rho(T)$ rising faster than $T^2$. (3) As \iToneT\ reaches the values of the `insulating master curve' (gray shaded; see Fig.~\ref{T1T-MIT}a,b), it forms a maximum that is most pronounced close to the Mott MIT ($x\rightarrow 0.16$ \cite{Pustogow2021-Landau,Pustogow2021-percolation}). This peak in \iToneT\ is well below $T_{\rm max}$, above which (4) $\rho(T)$ acquires a non-metallic slope. 
}
\label{transport-NMR}
\end{figure*}

The phase diagram of organic $\kappa$-\ET\ Mott systems assembled 
previously~\cite{Lefebvre2000,Limelette2003a,Kagawa2005,Komatsu1996,Kurosaki2005,Furukawa2015, Pustogow2018,Dressel2018,Furukawa2018,Pustogow2021-Landau,Pustogow2021-percolation} is
shown
in Fig.~\ref{intro}(b). One of the best studied materials 
with
a genuine Mott MIT is the frustrated triangular-lattice compound \Cu. It has been considered a quantum-spin-liquid candidate for decades~\cite{Shimizu2003,Shimizu2006,Yamashita2008,Yamashita2009,Balents2010}, whereas the recently discovered spin gap~\cite{Miksch2021} rather suggests a valence-bond-solid ground state~\cite{Pustogow2022,Pustogow2023}. So far, most attempts to modify the correlation strength focused on physical pressure~\cite{Komatsu1996,Kurosaki2005,Furukawa2015,Furukawa2018,Pustogow2021-percolation,Rosslhuber2021}, strain~\cite{Shimizu2011,Pustogow2023,Kawasugi2023}, or (partial) chemical substitution of the anions $X$~\cite{Faltermeier2007,Shimizu2016,Yoshida2019,Pustogow2021-TMTTF}. Substitution of the organic donor molecules provides another powerful tool to tune the Mott MIT in a controlled manner, which we deploy here in the alloys \STF\ \cite{Saito2018,Pustogow2021-Landau,Pustogow2021-percolation,Saito2021,Saito2021a,Yesil2023}, 
in the following abbreviated as \stf.

Single crystals were grown by electrochemical synthesis~\cite{Saito2018} and studied by four-point dc transport and NMR in the $T$ range 2--200~K. Non-selective saturation-recovery $^1$H NMR measurements of the spin-lattice relaxation rate \iTone\ were performed with out-of-plane magnetic field. Due to excessively long $T_1$ for $x=1$ ($10^4$~s at lowest $T$), we carried out additional $^{13}$C measurements.

Figure~\ref{T1T-MIT} displays our $T$-dependent NMR results for all substitutions. Notably, in the Mott-insulating state ($x< 0.16$ in Fig.~\ref{T1T-MIT}a), all \iTone\ values collapse on top of each other. This `insulating master curve' (highlighted in Fig.~\ref{T1T-MIT}b in gray) arises from the relaxation of local moments and comes along with a pronounced $T$ dependence of \iToneT , decreasing with increasing $T$. 
Overall, our measurements in the insulating state agree well with previous NMR studies using $^1$H and $^{13}$C nuclei~\cite{Shimizu2003,Kurosaki2005,Shimizu2006}, where $^1T_1$ is about 80 times longer than $^{13}T_1$~\cite{Saito2018,Pustogow2020HgCl,Pustogow2022}. The sample with $x = 0.16$ (shown only in Fig.~\ref{T1T-MIT}c to keep panels (a) and (b) clear) exhibits multi-exponential relaxation due to phase coexistence of insulating and metallic contributions, which is 
in accord with our dielectric results in Ref.~\cite{Pustogow2021-percolation} and will be the focus of a separate work.
Once coherent metallic transport is established for $x \!\geq\! 0.19$, we observe $T$-independent \iToneT\ at $T<T_{\rm FL}$, in accord with the Korringa law \iToneT $\propto K^2$~\cite{Alloul2014}.
Here, $K$ is the NMR Knight shift,
which is proportional to the uniform spin susceptibility $\chi(q=0)$.

Comparing the NMR data with the effective mass enhancement from recent optical and specific heat work on these systems~\cite{Pustogow2021-Landau,Yesil2023}, we find \iToneT $\propto (m^{\star}/m)^2$ within the FL regime. While it is well-known that the QP effective mass affects transport properties---relationships with the $A$-coefficient (Kadowaki--Woods~\cite{Kadowaki1986,Jacko2009}) and, lately, also with $\rho_0$~\cite{Efimova2025} have been discovered---a scaling with the NMR relaxation rate has not been reported previously. 
If we assume that the Korringa law holds throughout the metallic regime,
our findings indicate that the Wilson ratio ($R_{\mathrm{W}} \propto \chi/C$ with the specific heat coefficient $C \propto m^{\star}/m$) remains rather unaffected by varying correlation strength in the inspected region of the phase diagram. Nevertheless, firm conclusions on the correlation dependence of spin fluctuations and the corresponding Landau parameter require dedicated experimental studies (i) closer to the MIT and (ii) with higher resolution in substitution and/or pressure.

\begin{figure*}
\centering
 \includegraphics[width=2.05\columnwidth]{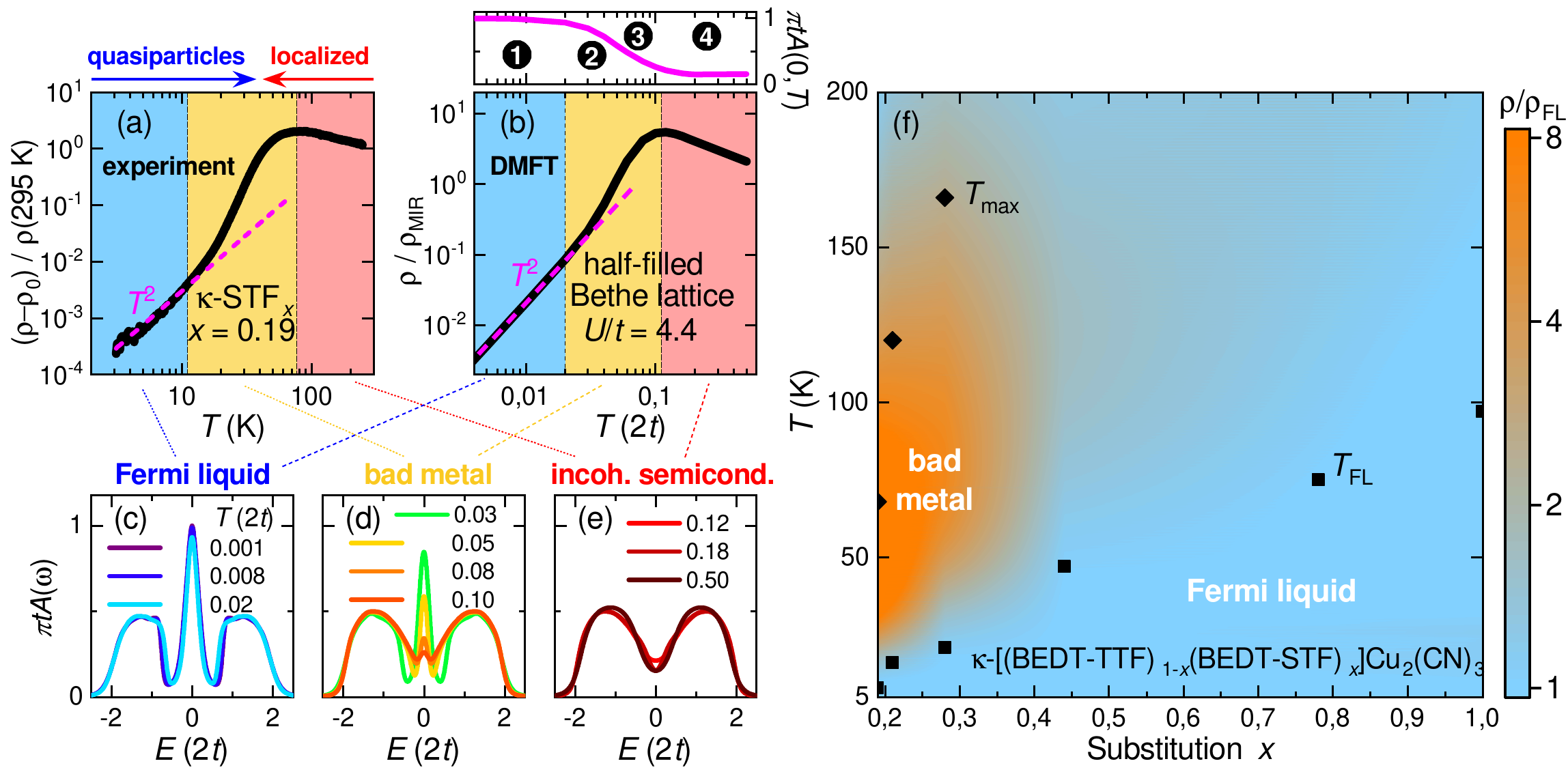}
\caption{
The (a) experimental resistivity is accurately reproduced by (b) DMFT. (c) The FL regime with 
$\rho\propto T^2$  has a pronounced, almost $T$-independent QP peak in the DMFT spectral function $A(\omega,T)$. (d) The increase of $\rho$ with $T$ well above $\rho_{\rm FL}$ and beyond $\rho_{\mathrm{MIR}}$ in the bad-metal regime is concomitant with a rapid reduction of the QP peak. The $T$-dependent height of the QP peak $A(0,T)$ is displayed on top of panel (b). (e) Semiconducting behavior above $T_{\rm max}$ is found when coherent QP have vanished completely. (f) The contour plot of $\rho(T)/\rho_{\rm FL}(T)$ measured on \stf\ illustrates the rise of $\rho$ faster than $T^2$ in the bad metal (orange). $\rho_{\rm FL}(T)$ corresponds to $T^2$ fits (black dashed lines) in Fig.~\ref{transport-NMR}(f-j); $x=0.78$ and 1.00 data from Ref.~\cite{Pustogow2021-Landau}.
}
\label{DMFT}
\end{figure*}

Next, we compare \iToneT\ to $\rho(T)$ measured on the \textit{same} single crystals (for details, see End Matter). Figures~\ref{transport-NMR}(a-e) show $\rho\propto T^2$ behavior extending up to $T_{\rm FL}$, followed by a steeper increase at higher $T$. The maxima of $\rho$ at $T_{\rm max}$ (Fig.~\ref{transport-NMR}f-j) are in good agreement (
several K sample-to-sample variation for nominally same $x$) with previous studies~\cite{Pustogow2021-Landau,Saito2021}. Crucially, for all studied metallic substitutions, $0.19\leq x\leq 1.00$, the $T$-independent \iToneT\ in Fig.~\ref{transport-NMR}(k-o) occurs exclusively when $\rho\propto T^2$ holds in the FL regime $T<T_{\rm FL}$ (region (1) in Fig.~\ref{intro}b).

This brings us to the main result of this work. We find that \iToneT\ acquires a $T$ dependence above $T_{\rm FL}$ and begins to increase simultaneously as $\rho$ rises beyond $\rho_{\rm FL}(T)$. Note that $\rho_{\rm MIR}=hd/e^2=4$~m$\Omega$ cm is surpassed at a temperature above $T_{\rm FL}$ but well below $T_{\rm max}$~\cite{Pustogow2021-Landau} (see Fig.~\ref{transport-NMR}(h) and End Matter)~\cite{Pustogow2021-Landau}. 

The maxima of $\rho$ and \iToneT\ become smaller with larger substitution as correlations diminish. We highlight the enhancement of $\rho$ beyond $\rho_{\rm FL}$ by orange in Fig.~\ref{transport-NMR}(f-j) and in the false-color contour plot of $\rho(T)/\rho_{\rm FL}(T)$ in Fig.~\ref{DMFT}(f).
Notably, the peak of \iToneT\ occurs at a lower $T$ than the maximum of $\rho$, thus dividing the $\rho \neq \rho_{\rm FL}$ metallic regime between $T_{\rm FL}$ and $T_{\rm max}$ into two regions indicated as (2) and (3) in Fig.~\ref{transport-NMR}(f,k). We interpret the experimental findings as follows: (1) FL state with well defined QP; (2) `resilient' QPs above $T_{\rm FL}$ \cite{Deng2013} are gradually destroyed,  yielding a steep rise of $\rho(T)>\rho_{\rm FL}(T)$ and an increase of \iToneT\ until it reaches the value of the relaxation of local moments (see Fig.~\ref{T1T-MIT}(a,b) and Fig.~\ref{T1T-model} in End Matter); (3) the NMR response is dominated by local moments (\iToneT\ decreasing with $T$) while $\rho$ is still metallic; the slower rise of $\rho$ indicates the demise of QPs, and $d\rho / dT>0$ with $\rho > \rho_{\rm MIR}$ the bad-metal character;
(4) both $\rho(T)$ and \iToneT\ exhibit the non-metallic response of a thermally activated semiconductor.
We thus find that $T_{\rm FL}$ is not the point where QP vanish, 
but where they gradually lose coherence.
Hence, regime (2) is still dominated by QP, yet their contribution to transport and NMR relaxation diminishes with increasing $T$.
We further note that $T_{\rm FL}$ and $T_{\rm MIR}$ in the present half-filled system are closer to each other than in a previous theoretical study of a doped Mott insulator \cite{Deng2013}. Nevertheless, they are distinct scales that can be distinguished experimentally as well as theoretically (see End Matter).

To obtain insight into the QP spectrum, we performed state-of-the-art dynamical mean-field theory (DMFT) \cite{Georges1996} computations for a half-filled Bethe lattice. Figures~\ref{DMFT}(a,b) show excellent agreement between $\rho(T)$ measured in the most strongly correlated sample ($x = 0.19$) and $\rho(T)$ calculated with DMFT for $U/t=4.4$, where $2t$ is the half-bandwidth of the semicircular lattice density of states. The associated spectral function $A(\omega)$ displayed in panels (c-e) remains almost unchanged in the FL regime and acquires a strong $T$ dependence above $T_{\rm FL}$ as the QP peak 
diminishes with increasing $T$ (see $A(0,T)$ on top of Fig.~\ref{DMFT}b). For $T>T_{\rm max}$, the QP excitations have vanished entirely, leaving a thermally filled pseudogap. 
This is consistent with the phase diagram in Fig.~\ref{intro}(b) and the experimental data in Fig.~\ref{transport-NMR}(f-j), as well as previous studies~\cite{Limelette2003,Georges_iscom_2004}.
The bad-metal regime is, thus, the crossover region in which the QP peak disappears 
and $\rho$ steeply rises as the conducting QP states are destroyed.

\begin{figure}[t]
\centering
\includegraphics[trim={0.4cm 1.3cm 0 0.9cm},clip,width=1\columnwidth]{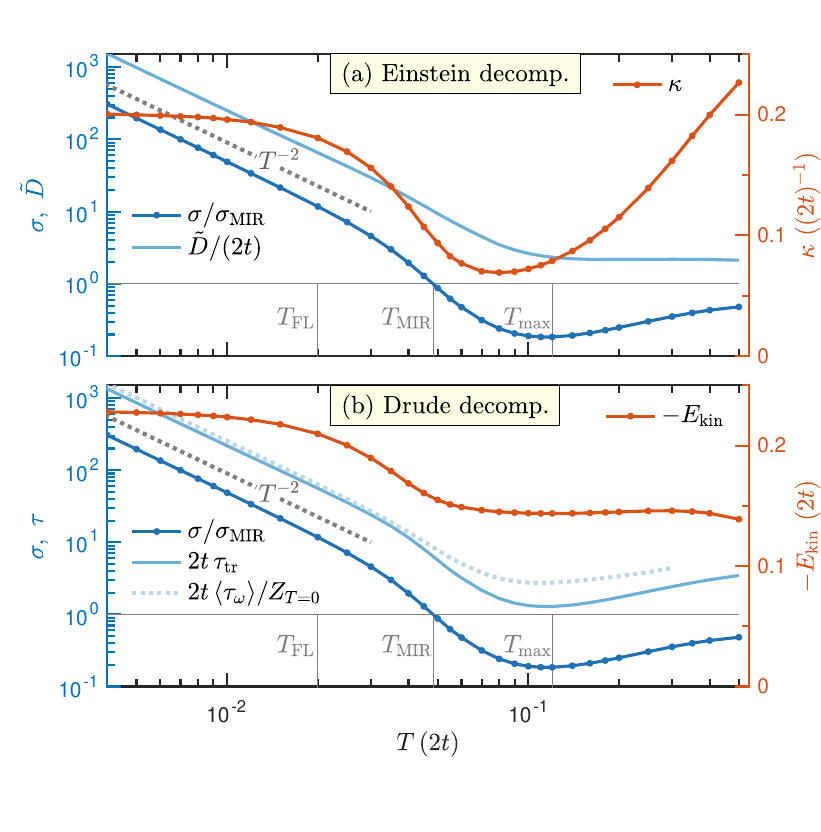}
\caption{Decompositions of the DMFT conductivity $\sigma$ (inverse of $\rho$ from Fig.~\ref{DMFT}b).
(a) The Einstein relation $\sigma = D \kappa$, or
$\sigma/\sigma_{\rm MIR} = \tilde{D} \kappa$,
yields the diffusion constant $D$ from $\sigma$ and the compressibility $\kappa = \partial n / \partial \mu$.
(b) The Drude-like relation $\sigma/\sigma_{\rm MIR} = \tau_{\rm tr} E_{\rm kin}$ yields the transport scattering time $\tau_{\rm tr}$ from $\sigma$ and the kinetic energy $E_{\rm kin}$.
In the FL regime, $\tau_{\rm tr}$ is comparable to the QP scattering time $\langle {\tau_\omega} \rangle/Z$ (see End Matter).
}
\label{DMFT_sigma_decomposition}
\end{figure}

Next, we decompose (Fig.~\ref{DMFT_sigma_decomposition}) 
the DMFT result for $\sigma=1/\rho$ in two ways, using the Einstein relation $\sigma = D \kappa$ and a Drude-like expression  
$\sigma/\sigma_{\rm MIR} = \tau_{\rm tr} E_{\rm kin}$ 
(where the kinetic energy $E_{\rm kin}$ can be interpreted as a carrier density, from the optical sum-rule) \cite{Perepelitsky2016}. 
In the FL regime, the $T^{-2}$ dependence of $\sigma$ comes from the diffusion constant $D$ and the transport scattering time $\tau_{\rm tr}$ in the respective pictures, while the electronic compressibility $\kappa$ and the kinetic energy $E_{\rm kin}$ are constant.

For $T>T_{\rm FL}$, the Drude-like decomposition involves a decrease of $E_{\rm kin}$ and of $\tau_{\rm tr}$.
Whereas $\sigma$ falls below $\sigma_{\rm MIR}$, $\tau_{\rm tr}$ stays just above $\hbar/(2t)$,
corresponding to a scattering event on every site.
In the End Matter, we also use the Kubo formula to show that the Drude result in terms of the QP scattering rate 
acquires sizable corrections for $T>T_{\rm FL}$, which are amplified in magnitude by a small QP weight $Z$.

By contrast, the Einstein decomposition provides a clearer picture: $D$ is rather featureless in the considered range of $T$, merely leveling off around $T_{\rm max}$. The strong decrease of $\sigma$ with a local minimum is entirely driven by $\kappa$, which displays a local minimum at which it is reduced by more than a factor of two, corresponding to an almost incompressible state in the bad-metal regime from $T_{\rm MIR}$ to $T_{\rm max}$. 
This effect of $\kappa$ on $\sigma$ is in perfect agreement with the insulating character seen in the single-particle spectrum as well as in NMR. 

Note, however, that dc transport alone cannot directly probe the evolution of the QP band-structure, calling for energy-resolved techniques. 
Indeed, spectroscopic experiments on oxides and organics revealed that, even in a regime with apparent non-FL behavior, charge transport is still based on $resilient$ QP~\cite{Deng2013,Pustogow2021-Landau} with a $T$-dependent QP weight $Z$ as seen in various transition-metal oxides~\cite{Deng2014,Hunter2023}. In the case of \sro, the increase of $Z(T)$ with $T$ agrees with the $T$ dependence of $\chi(T)$ determined by the NMR Knight shift $K$~\cite{Chronister2022}.
A $T$ dependence of $m^\star$ was also reported in doped \sto\ \cite{Collignon2020,Kumar2024}, but there $Z(T)$ reduces with higher $T$ in line with an increase of $\rho(T)$ beyond FL-type $AT^2$. 
In FeSe an, orbital-selective spectral weight loss was revealed by ARPES~\cite{Yi2015}, while Hall-effect measurements in the cuprates suggested that a $T$-dependent carrier density $N(T)$ could explain the linear-$T$ dependence of $\rho$~\cite{Barisic2019,Klebel-Knobloch2023}.

In conclusion, we studied the bad metal arising from a FL with well-defined QP upon increasing $T$, for varying correlation strengths in the vicinity of the Mott MIT in \STF. In the FL regime, $T<T_{\rm FL}$,  Korringa law applies, and we find that \iToneT\ scales with $(m^\star/m)^2$ obtained from previous optical studies~\cite{Pustogow2021-Landau}. When $\rho$ exceeds the FL-type $T^2$ behavior and $\rho_{\rm MIR}$ in the bad metal regime, \iToneT\ also acquires a $T$ dependence. Already before the resistivity maximum is reached, the NMR relaxation is essentially identical to the response of local moments in a thermally activated Mott insulator. 
Our study thus demonstrates that bad-metal behavior near the Mott MIT is not due to unrealistically large scattering rates that go way beyond the MIR limit. Instead, it results from the gradual destruction of QP excitations, that vanish entirely at $T>T_{\rm max}$ yielding insulating behavior. Our DMFT calculations clearly visualize this in the QP spectrum and reproduce $\rho(T)$ very well near the Mott MIT. Our work highlights the necessity to monitor $T$-dependent changes of the QP spectrum (weight and dispersion) in systems exhibiting apparent non-FL transport. 

Our study also has broader implications, in particular in connection to moir\'e materials. 
There, the emergence of local moments upon heating a metallic state, 
signaled by the Pomeranchuk effect~\cite{Rozen2021,Saito_TBLG_2021} 
is often interpreted in terms of a two-fluid Kondo-like picture. 
Our experimental and theoretical results indicate that QPs get gradually converted into 
local moments even when a single species of electrons is present, as is the case of the BEDT organics.
A breakthrough development on moir\'e materials has been the ability to independently 
measure transport and electronic compressibility via gating~\cite{Tomarken2019,Rozen2021,Saito_TBLG_2021}. 
Such a direct measurement of the electronic compressibility in BEDT salts would be highly desirable, 
allowing in particular to test our Einstein relation-based prediction relating the maximum of resistivity 
(also observed in some moir\'e materials close to a Mott state \cite{Wang2020,Ghiotto2021,Li2021,Zang2022}) 
to a minimum of the compressibility. 

\acknowledgments  We thank M.~Dressel, S.~Fratini and V.~Dobrosavljevic for useful comments and discussions,
Seung-Sup~B.~Lee for discussions about anomalous contributions,
and M.~Gievers for providing quantum Monte Carlo data.
The DMFT equations were solved with the numerical renormalization group (NRG) \cite{Bulla2008} using a symmetric improved estimator for the self-energy \cite{Kugler2022};
the implementation uses the MuNRG package~\cite{SSLee2016,SSLee2017,SSLee2021} built on top of the QSpace tensor library~\cite{Weichselbaum2012a,Weichselbaum2020,Weichselbaum2024}.
We acknowledge funding by the Czech Science Foundation via research project GA\v{C}R 23-06810O.
F.B.K.\ acknowledges funding from the Ministerium f\"ur Kultur und Wissenschaft des Landes Nordrhein-Westfalen (NRW-R\"uckkehrprogramm).
The Flatiron Institute is a division of the Simons Foundation.

\bibliography{Literatur}
\bibstyle{unsrt}

\clearpage

\onecolumngrid
\section{End Matter}
\twocolumngrid

\textit{Details of Transport Experiments}---%
Subsequent to our NMR studies, $\rho$ (Fig.~\ref{transport-NMR}) was measured for $x\leq 0.44$ on the same single crystals in the same PPMS cryostat 
with in-plane electric field parallel to the $c$-axis. This has the crucial advantage that we can directly assign the transport and NMR data at the very same trajectory in the phase diagram without external pressure. Usually, sample-to-sample variations and different pressure conditions at different cooldowns do not allow one to reproduce the same positions in the $T$-$p$ phase diagram. Even if the same sample was measured in the same pressure cell first by NMR and then by transport, it would be impossible to realize the same conditions in two different pressure runs because issues like pressure loss upon cooling (as in typical CuBe oil pressure cells) cause considerable error bars.

For the quasi-2D organic materials under study, an inter-layer distance of $d=16\AA$ yields $\rho_{\rm MIR}=hd/e^2=4$~m$\Omega$ cm. Note, however, that in the present study the sample geometry could not be determined with sufficient precision to allow for a quantitative comparison of the measured $\rho$   with $\rho_{\rm MIR}$. In previously published transport data on \STF\ \cite{Pustogow2021-Landau} $\rho_{\rm MIR}$ is reached at a temperature not far above $T_{\rm FL}$, well below $T_{\rm max}$, as indicated in Fig.~\ref{transport-NMR}(h) for $x=0.28$ that was compared to the data from Ref.~\cite{Pustogow2021-Landau}.

\begin{figure}
\centering
 \includegraphics[width=1\columnwidth]{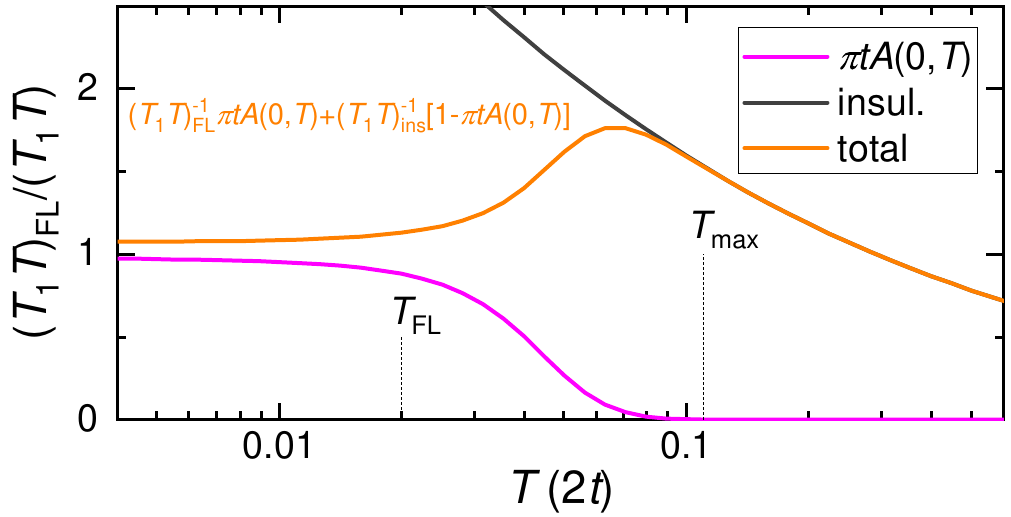}
\caption{In the low-$T$, FL regime, \iToneT\ is rather constant. At $T>T_{\rm FL}$, the reduction of QP excitations (approx. here by $A(0,T)$) yields a reduction of the metallic \iToneT. At the same time, the insulating component gains weight (prop. to $1-A(0,T)$), yielding an increase of the total \iToneT\ as the insulating component exceeds the metallic one. Once the insulating value is reached, it decreases with temperature.
}
\label{T1T-model}
\end{figure}

\textit{Formation of the peak in \iToneT }---%
The gradual reduction of QP excitations in the bad-metal crossover yields complex behavior in the spin-lattice relaxation rate. As the metallic component reduces below the FL value, the overall \iToneT\ is enhanced because the faster relaxing insulating component gains weight. As the QP excitations have significantly reduced -- but not fully disappeared, NMR relaxation becomes dominated by local moments. Thus, the total \iToneT\ acquires an insulating-like response and reduces with temperature, resulting in a local maximum between $T_{\rm FL}$ and $T_{\rm max}$, as shown in Fig.~\ref{T1T-model}.

\textit{Decomposing the DMFT resistivity}---%
We compute the conductivity from the Kubo formula,
\begin{align}
\sigma & = \int d\omega \, (-f'_{\omega}) \mathcal{T}_{\omega}
, \quad
\mathcal{T}_{\omega} =
2\pi \int \frac{d^d k}{(2\pi)^d} v_{\vec{k}}^2 A_{\vec{k},\omega}^2
,
\label{eq:Kubo}
\end{align}
neglecting vertex corrections.
Here, $f'_\omega$ is the derivative of the Fermi function with respect to frequency, $d$ is the dimensionality, and the momentum integral extends over the first Brillouin zone.
We use atomic units where $k_{\text{B}}=\hbar=e=1$;
$v_{\vec{k}} = \partial_{k_\alpha} \epsilon_{\vec{k}}$ is the band velocity in a given direction $\alpha \in \{ x, y, z \}$, and $A_{\vec{k}\omega}$ is the spectral function. 

In DMFT, the self-energy $\Sigma_\omega$ is local, and the spectral function depends on $\vec{k}$ only via $\epsilon_{\vec{k}}$, 
\begin{align}
A_{\epsilon_{\vec{k}},\omega}
=
-\frac{1}{\pi} \mathrm{Im} \frac{1}{\omega+\mu-\epsilon_{\vec{k}}-\Sigma_\omega}
,
\end{align}
with the chemical potential $\mu$. We can then write $\mathcal{T}_{\omega}$ as 
\begin{align}
\mathcal{T}_{\omega} & =
2 \pi \int d\epsilon \, \Phi_\epsilon A_{\epsilon,\omega}^2
, 
\ \
\Phi_\epsilon 
=
\int \frac{d^d k}{(2\pi)^d} 
v_{\vec{k}}^2
\delta(\epsilon - \epsilon_{\vec{k}})
.
\end{align}

Following Refs.~\citenum{Berthod2013,Georges2021},
we define
the (renormalized) Fermi energy $E^{\mathrm{F}}=\mu - \mathrm{Re}\Sigma_0$, the generalized QP weight $Z_{\omega}$ 
via $1 - 1/Z_{\omega} = [\mathrm{Re}\Sigma_\omega - \mathrm{Re}\Sigma_0]/\omega$,
and the generalized scattering rate 
$\Gamma_{\omega} = 1/\tau_\omega = -\mathrm{Im}\Sigma_\omega$.
For $\omega\to 0$, one recovers the standard QP weight $Z_0$ and scattering rate $\Gamma_0$. Via $\Sigma_\omega$, all of the above quantities implicitly depend on $T$.
Inserting the resulting expression for $A_{\epsilon,\omega}$ into $\mathcal{T}_\omega$ and
substituting \cite{Georges2021} 
$\epsilon = E^{\mathrm{F}}+\omega/Z_\omega + \Gamma_\omega y$
,
we obtain
\begin{align}
\mathcal{T}_{\omega}
& = 
2\pi \int_{-D}^D \mathrm{d} \epsilon \, \Phi_\epsilon
\bigg[
\frac{\Gamma_\omega / \pi}{(E^{\mathrm{F}}+\omega/Z_\omega-\epsilon)^2+\Gamma_\omega^2}
\bigg]^2
\nonumber \\
& =
\frac{2}{\pi\Gamma_\omega}
\int_{D^-_\omega}^{D^+_\omega} \mathrm{d} y \, \Phi_{
E^{\mathrm{F}}+ \frac{\omega}{Z_\omega}+\Gamma_\omega y 
}
\bigg[ \frac{1}{1+y^2} \bigg]^2
,
\end{align}
where we assumed that $\Phi_\epsilon$ has a finite bandwidth $-D$ to $D$ and used
$D^\pm_\omega = (\pm D-E^{\mathrm{F}}-\omega/Z_\omega)/\Gamma_\omega$.
Typically, $\Phi$ varies slowly in the range set by the relevant $\omega$ and $y$. Hence, we may expand $\Phi$ up to second order:
\begin{align}
\mathcal{T}_{\omega} \Gamma_\omega & =
\Phi_{E^{\mathrm{F}}} g^{(0)}_\omega + 
\Phi'_{E^{\mathrm{F}}} \bigg( \frac{\omega}{Z_\omega} g^{(0)}_\omega + \Gamma_\omega g^{(1)}_\omega \bigg) 
\nonumber
\\
& \ + 
\frac{\Phi''_{E^{\mathrm{F}}}}{2} \bigg( \frac{\omega^2}{Z_\omega^2} g^{(0)}_\omega + \frac{2\omega\Gamma_\omega}{Z} g^{(1)}_\omega + \Gamma_\omega^2 g^{(2)}_\omega \bigg)
.
\end{align}
Here, we defined $g^{(n)}_\omega = h_n(D^+_\omega) - h_n(D^-_\omega)$ and 
$\pi h_n(z) = 2 \int^z d y \, y^n (1+y^2)^{-2}$.
We thus have
$\pi h_0(z) = \arctan(z) + \frac{z}{1+z^2}$,
$\pi h_1(z) = - \frac{1}{1+z^2}$,
$\pi h_2(z) = \arctan(z) - \frac{z}{1+z^2}$,
and consequently $\lim_{|D|/\Gamma_\omega \to \infty} \{ g_0, g_1, g_2 \} = \{ 1, 0, 1 \}$.
Note that the second-order term and our account of bandwidth effects through $g^{(n)}_\omega$ go beyond Ref.~\citenum{Georges2021}.

\begin{figure}[t]
\centering
\includegraphics[trim={0.5cm 0 0.2cm 0},clip,width=1\columnwidth]{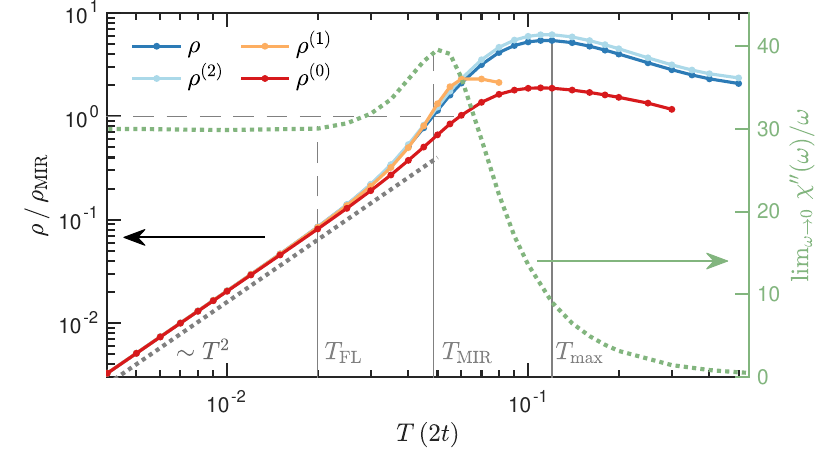}
\caption{Left axis: DMFT results for $\rho = 1 / \sigma$ as in Fig.~\ref{DMFT} together with the approximations of Eqs.~\eqref{eq:sigma2}, \eqref{eq:sigma1}, and \eqref{eq:sigma0}. We use the Bethe lattice, where $\Phi_\epsilon \!=\! \Phi_0 [1 - \epsilon^2/(2t)^2]^{3/2}$ and $\rho_{\mathrm{MIR}} = 2t/\Phi_0$ \cite{Deng2013}. 
We find that $T \!>\! T_{\rm FL}$ can also be identified with $Z|\mathrm{Im}\Sigma(0)| \!>\! T$
and $\rho^{(0)} \!>\! \rho_{\rm MIR}$ with 
$\langle\tau_\omega\rangle^{-1} \!>\! 2t$.
Right axis: DMFT results for the local retarded spin susceptibility $\chi$. 
The normalization is such that $\lim_{\omega\to 0}\chi''(\omega)/\omega \!=\! 1/(2\pi^2 t^2)$ at $U \!=\! 0$.
}
\label{DMFT_expansion}
\end{figure}

We now consider particle-hole symmetry
where $E^F=0$ and $\Phi(\epsilon)=\Phi(-\epsilon)$.
The result for $\sigma$ is then
\begin{align}
\sigma = 
\bigg\langle \tau_\omega
\bigg(
\Phi_0 g^{(0)}_\omega + \frac{\Phi''_0}{2} \bigg[\frac{\omega^2}{Z_\omega^2} g^{(0)}_\omega
\!+\! 
\frac{2\omega\Gamma_\omega}{Z_\omega} g^{(1)}_\omega 
\!+\! 
\Gamma_\omega^2 g^{(2)}_\omega
\bigg]
\bigg)
\bigg\rangle
,
\label{eq:sigma2}
\end{align}
where $\langle \cdot \rangle \equiv \int d\omega \cdot (-f'_\omega)$.
We call this expression $\sigma^{(2)}$ as it follows from Eq.~\eqref{eq:Kubo} by a second-order expansion of $\Phi_\epsilon$.
We also consider two more simplified expressions, in which we use the wide-band limit $\{ g_0, g_1, g_2 \} = \{ 1, 0, 1 \}$: The simplest, Drude-like result is the first term of Eq.~\eqref{eq:sigma2},
\begin{align}
\sigma^{(0)} = 
\Phi_0 \langle \tau_\omega \rangle
.
\label{eq:sigma0}
\end{align}
In the FL regime ($|\omega|,T \!\ll\! T_{\mathrm{FL}}$), one has
$-\mathrm{Im}\Sigma_\omega
\!=\! C(\omega^2 + \pi^2 T^2)$.
By standard integration,
we then find 
$\langle \tau_\omega \rangle
=
1/(12 C T^2)
=
\tau_0 \pi^2 /12$
and thus $\rho^{(0)} = 1/\sigma^{(0)} = A T^2$ with 
$A = 12 C / \Phi_0$
\cite{Kugler2025,LaBollita2025}.
(There is a factor 2 difference to these references since we, following Ref.~\cite{Deng2013}, did not include the spin sum in $\Phi$.)

\begin{figure}[t]
\centering
\includegraphics[trim={0.9cm 0 0.9cm 0.4cm},clip,width=1\columnwidth]{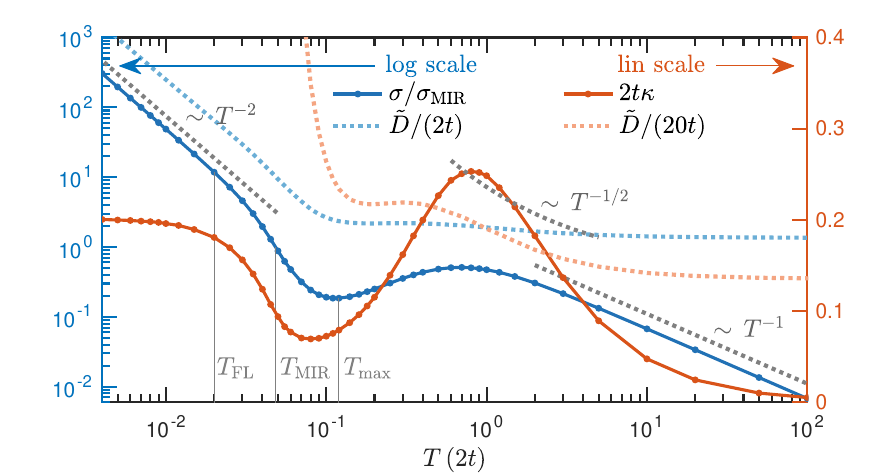}
\caption{DMFT results for $\sigma=1/\rho$ decomposed according to the Einstein relation,
as in Fig.~\ref{DMFT_sigma_decomposition} but for a larger $T$ range.
}
\label{DMFT_sigma_kappa}
\end{figure}


For the leading correction to $\sigma^{(0)}$ in $T/T_{\mathrm{FL}}$, we may estimate $1/Z_\omega \sim D/T_{\mathrm{FL}}$ and $\omega/Z_\omega \sim D T/T_{\mathrm{FL}}$ for $|\omega| \lesssim T$.
By contrast, $\Gamma_\omega \sim D (T/T_{\mathrm{FL}})^2$ scales quadratically with $T/T_{\mathrm{FL}}$ in that regime.
Hence, we get
\begin{align}
\sigma^{(1)}
& =
\bigg\langle \tau_\omega
\bigg(
\Phi_0  \!+\! \frac{\Phi''_0}{2} \frac{\omega^2}{Z_\omega^2} 
\bigg)
\bigg\rangle
\sim
\sigma^{(0)}
\bigg[ 1 \!+\! 
\frac{\Phi''_0 D^2}{2 \Phi_0}
\bigg( \frac{T}{T_{\mathrm{FL}}} \bigg)^2
\bigg] 
.
\label{eq:sigma1}
\end{align}
Typically $\Phi''_0<0$, so that the correction to $\sigma^{(0)}$ is negative and that to $\rho^{(0)} = 1/\sigma^{(0)}$ positive.
Note that, while the Drude-like $\sigma^{(0)}$ is independent of the (generalized) QP weight $Z_\omega$, the leading correction $\sigma^{(1)}$ is enhanced by a factor $1/Z_\omega^2$ for small $Z_\omega$.
This enhancement is enforced by the fact that $Z_0$ decreases with increasing $T$, but it would exist even without $T$-dependence in $Z_\omega$.

Figure~\ref{DMFT_expansion} (left axis) shows DMFT results for $\rho/\rho_{\mathrm{MIR}}$, where we compare the full result from Eq.~\eqref{eq:Kubo} to the three approximations outlined above.
In the FL regime, the Drude-like expression Eq.~\eqref{eq:sigma0}, yielding $\rho^{(0)} = AT^2$, is sufficient. For $T > T_{\mathrm{FL}}$, corrections to $\rho^{(0)}$ become noticeable. They are indeed positive and well captured by $\rho^{(1)}$ of Eq.~\eqref{eq:sigma1}. For larger $T$, where 
$\langle\tau_\omega\rangle^{-1} > 2t$, effects of the finite bandwidth enter, rendering Eq.~\eqref{eq:sigma1} inappropriate. However, the quadratic expansion of $\Phi_\epsilon$, leading to Eq.~\eqref{eq:sigma2}, remains appropriate for all $T$ considered in Fig.~\ref{DMFT_expansion}, as $\rho^{(2)}$ reproduces $\rho$ up to a minor shift.

The right axis of Fig.~\ref{DMFT_expansion} shows a DMFT proxy for \iToneT, 
$\lim_{\omega\to 0}\chi''(\omega)/\omega$,
where $\chi''$ is the imaginary part of the local retarded spin susceptibility.
In agreement with experiment (Fig.~\ref{transport-NMR}), 
$\lim_{\omega\to 0}\chi''(\omega)/\omega$
is constant below $T_{\rm FL}$, decays for large $T$ and, importantly, has a maximum in the bad-metal crossover between $T_{\rm FL}$ and $T_{\rm max}$. Here, the peak is found very close to $T_{\rm MIR}$.
There is a subtlety in the calculation: at finite $T$, the Wilson chain in NRG is effectively cut off \cite{Weichselbaum2012}, rendering the system finite with a non-negligible anomalous part 
\cite{Watzenboeck2022,Kugler2021}
in $\lim_{\omega\to 0}\chi''(\omega)/\omega$.
It is given by $C \delta(\omega)/T$ with
$C = \sum_{nm}^{E_n=E_m} \rho_n |S^z_{nm}|^2$, where $n,m$ are eigenstates, $\rho$ is the density matrix, and $S^z$ the spin operator. To represent the continuum system, $\delta(\omega)$ must be broadened, acquiring a finite slope. We choose the one broadening parameter manually so that $\lim_{\omega\to 0}\chi''(\omega)/\omega$ matches with quantum Monte Carlo data \cite{Watzenboeck2022} available at higher $T$.

Finally, Fig.~\ref{DMFT_sigma_kappa} shows DMFT results for $\sigma=1/\rho$ decomposed according to the Einstein relation $\sigma = D \kappa$, or $\sigma/\sigma_{\rm MIR}=\tilde{D}\kappa$, as in Fig.~\ref{DMFT_sigma_decomposition}
but for larger $T$---%
even if not accessible in \stf.
In the high-$T$ limit,
$D$ approaches a constant while $\sigma$ and $\kappa$ decay as $T^{-1}$.
Additionally,
approaching the high-$T$ thermal state, $D$ is somewhat consistent with a $T^{-1/2}$ decay \cite{Eom2025}.

\end{document}